\documentclass[aps,preprint,showpacs]{revtex4}
\usepackage{amssymb}
\usepackage{graphicx}
\usepackage{bm}
\usepackage{amsmath}

\begin{document}
\pacs{71.27.+a, 71.10.Fd, 71.20.Be}
\title{The intermediate pressure phases of cerium studied by LDA+Gutzwiller method}
\author{Ming-Feng Tian$^{1,2}$, Xiaoyu Deng$^{2,3}$, Zhong Fang$^2$, Xi Dai$^{2}$}
 \affiliation{$^1$Institute of Applied Physics and Computational Mathematics, P. O. Box 8009, Beijing 100088, China\\
 $^2$Beijing National Laboratory for Condensed Matter Physics and Institute of Physics, Chinese Academy of Sciences, Beijing 100190, China\\
 $^3$Centre de Physicque Th$\acute{e}$orique, Ecole Polytechnique, CNRS, 91128 Palaiseau Cedex, France}

\date{\today}

\begin{abstract}
The thermodynamic stable phase of cerium metal in the intermediate pressure regime (5.0--13.0 GPa)
is studied in detail by the newly developed local-density approximation (LDA)+ Gutzwiller method, which can include the strong
 correlation effect among the 4\textit{f} electrons in cerium metal properly. Our numerical results
 show that the $\alpha''$ phase, which has the distorted body-centered-tetragonal structure, is the
 thermodynamic stable phase in the intermediate pressure regime and all the other phases including the
 $\alpha'$ phase ($\alpha$-U  structure), $\alpha$ phase (fcc structure), and bct phases are either
 metastable or unstable. Our results are quite consistent with the most recent experimental data.
\end{abstract}

\maketitle

\bigskip

\section{Introduction}

Due to the strong correlation effect, the 4\textit{f} electrons in Lanthanide metal usually  participate
very weakly into the chemical bonding, which makes these materials approximately \textit{s}-band
metal with close-packed crystal structure. A very important exception of this qualitative understanding
is the cerium metal, where the 4\textit{f} electrons participate in chemical bonding in the $\alpha$
fcc phase under ambient pressure. While the $\alpha$ fcc phase is quite close to the instability, an
isostructure phase transition happens by raising the temperature above 116 K, after which the crystal
structure remains unchanged while the volume expands by 16\% and the 4\textit{f} electrons become localized.\cite{Gshneidner_1962} Further numerical studies by implementing the first-principles methods
with dynamical mean-field theory\cite{DMFT,DMFT2}  show that the $\gamma$ phase may be stabilized by
the entropy.\cite{antoine}

Another mysterious phenomena in cerium is the intermediate pressure phase. At zero temperature, the
cerium metal forms the face-centered-cubic (fcc) structure for pressure below 5.0 GPa and body-centered-tetragonal
(bct) structure for pressure above 13.0 GPa. But the experimental results of the thermodynamic
stable phase in the intermediate pressure region between 5.0 and 13.0 GPa are still quite
controversial,\cite{Ellinger_1974, Guoliang_1995, Zhao_1997, Olsen_1985, McMahon_1997, Dmitriev_2004}
as will be discussed below in detail.

Ellinger and Zachariasen applied x-ray-diffraction studies on high-pressure cerium with a diamond-anvil cell,\cite{Ellinger_1974} and they reported that for pressure between 5.0 and 13.0 GPa, the orthorhombic
$\alpha'$-Ce phase with an $\alpha$-uranium type of structure is the thermodynamic stable phase. They
also found that the $\alpha''$ phase, which is monoclinic body centered with a deformed cubic face-centered
structure, is the thermodynamic metastable phase. The conclusion that $\alpha'$-Ce is the thermodynamic
stable phase between 5.0 and 13.0 GPa while the $\alpha''$ phase is metastable has been supported by some of
the follow-up experiments, i.e., Refs. \cite{Guoliang_1995} and \cite{Zhao_1997}, while another group of
experiments led to the opposite conclusion, which indicated that the $\alpha''$ rather than the $\alpha'$
phase is the thermodynamic stable phase in the intermediate pressure regime. Using diamond-anvil cell
and synchrotron radiation Olsen \textit{et al.} studied the high-pressure phase diagram of cerium up
to 46.0 GPa.\cite{Olsen_1985} They reported that the $\alpha''$ phase is the thermodynamic stable intermediate
pressure phase of Ce, and no evidence for the $\alpha'$ phase with $\alpha$-uranium structure was found.
This was the first experiment reporting the $\alpha''$ phase to be the thermodynamic stable intermediate
pressure phase for cerium. After that several other groups also reported similar experimental results
supporting the $\alpha''$ phase to be the thermodynamic stable phase of cerium for pressure between 5.0
and 13.0 GPa.\cite{McMahon_1997, Dmitriev_2004}

The first-principles calculation is a powerful theoretical tool to predict the ground-state phases of
solid. During the past two decades, many efforts have been made to reveal the thermodynamic stable phase
for cerium under intermediate pressure by first-principles calculations. The early results of the linear
muffin-tin orbital (LMTO) method found the $\alpha'$ phase to be the thermodynamic stable phase in
the intermediate pressure regime between the low-pressure fcc phase and high-pressure bct phase,\cite{Skriver_1985, Eriksson_1991} which was consistent with the early experiments. After that, S\"oderlind and Eriksson
\textit{et al.} applied the generalized-gradient approximation (GGA) based on the full-potential linear
muffin-tin orbital (FPLMTO) method to the same problem\cite{Soderlind_1995b} and found that $\alpha''$
phase was the thermodynamic stable phase only in a small pressure interval. Ravindran \textit{et al.}
have made systematic electronic structure and total-energy studies on Ce and did not find any thermodynamic
stable phase in the intermediate pressure regime between the low-pressure fcc phase to the high-pressure
bct phase.\cite{Ravindran_1998} According to their results, both the $\alpha'$ and $\alpha''$ phases
are metastable phases with the $\alpha''$ phase being lower in energy. After that, local-density
approximation (LDA)- or GGA-type calculations by other groups using the plane-wave method\cite{Richard_2001}
or the exact muffin-tin orbitals (EMTO) method\cite{Landa_2004} also got similar results, that both $\alpha'$
and $\alpha''$ phases are metastable phases and the thermodynamic stable phases of cerium are $\alpha$ phase
(low pressure) and bct phase (high pressure).

Although the 4\textit{f} electrons in the $\alpha$ phase are delocalized, the strong repulsive interaction
among them still modifies its electronic structure significantly. Due to the insufficient treatment
of the correlation effects, the bonding strength of the $\alpha$ cerium has been over estimated by
the LDA-type calculations, which leads to smaller volume and larger bulk modules compared with the
experimental data. In the present paper, we apply the newly developed LDA+Gutzwiller method, which
can satisfactorily treat the strong correlation effects in the 4\textit{f} shell, to determine the
thermodynamic stable phase of cerium under pressure. We first apply the above method to study the
ground-state properties of $\alpha$ cerium under the ambient pressure. Our results show that both the
volume and bulk modules are improved dramatically, which manifests the importance of the strong
correlation effect for the 4\textit{f} electrons in $\alpha$ cerium. Further we apply the same method
to study the intermediate pressure phases of cerium, and the results show that in the intermediate-pressure
region the $\alpha''$ phase is the thermodynamic stable phase and all other structures are
either metastable or unstable, which is quite consistent with the recent experiments.\cite{McMahon_1997, Dmitriev_2004}

The rest of the paper is organized as follows: A brief introduction of the LDA+Gutzwiller method are
given in Sec. \ref{sec:method}. In Sec. \ref{sec:results} we discuss the main results of our LDA+Gutzwiller
calculations with the comparison to the recent experimental results and LDA/GGA results. The summary
and conclusions are given in the last section.

\section{LDA+Gutzwiller method} \label{sec:method}
Gutzwiller first introduced the Gutzwiller variational approach to study the itinerant ferromagnetism
in systems with partially filled \textit{d} bands described by the Hubbard model.\cite{Gutzwiller_1960s}
Since then, the Gutzwiller variational approach has been widely applied to various strongly correlated systems.\cite{Bunemann, Brinkman, Vollhardt, FCZhang, Metzner} Recently, we developed a computational
method to incorporate LDA with the Gutzwiller variational approach, named the LDA+Gutzwiller method (simply called
LDA+G hereafter),\cite{GTWang, xydeng_2008, xydeng_2009} by successfully applying to a number of typical
correlated materials, the reliability and feasibility of this method have been demonstrated. In the
following we present the method briefly; please refer to our previous paper\cite{xydeng_2009} for more
details.

Similar to LDA+U and LDA+DMFT methods, in LDA+G the LDA Hamiltonian, which can be extracted from
the first-principles calculation, is implemented by a Hubbard-like local Coulomb interaction, which
is not adequately treated within LDA. The effect of this local Coulomb interaction can thus be considered
within the Gutzwiller variational approach. The Hamiltonian can be usually expressed as

\begin{equation} \label{eq:H}
H =H_{LDA}+H_{int}-H_{DC},
\end{equation}
with
\begin{equation} \label{eq:Hint}
H_{int}=\sum_{i, \alpha ,\beta(\alpha \neq \beta)}{
\mathcal{U}}_{i}^{\alpha ,\beta}\hat{n}_{i\alpha }\hat{n}_{i\beta},
\end{equation}

where $H_{LDA}$ is the LDA part of the Hamiltonian extracted from the standard LDA calculation, $H_{int}$
is the on-site interaction term, where $\alpha$ and $\beta$ are combined spin-orbit indices of localized
basis $ \{\phi _{i, \alpha }\}$ on site $i$, among which the local Hubbard interaction is implemented,
$\alpha=1,\ldots,2N$ ($N$ is the orbital number, e.g., $N=7$ for $f$ electrons). $H_{DC}$ is the double
counting term representing the average orbital independent interaction energy already included by LDA.
Without the $H_{int}$ term, the ground state can be exactly given by the Kohn-Sham uncorrelated wave
function (KSWF) $\vert \Psi_0\rangle $, which is a single Slater determinant made from the single-particle
wave functions. However, with the increment of the interaction strength, the KSWF is no longer a good
approximation because it gives too much weighting factor for those energetically unfavorable configurations.
In order to give a better description of the ground state, the weighting factor of those unfavorable
configurations should be suppressed, which is the main idea of Gutzwiller wave functions (GWFs)
$\vert \Psi_G \rangle $. A GWF is constructed by a many-particle projection operator acting on the uncorrelated
KSWF, which reads

\begin{equation} \label{eq:psi}
\vert \Psi_G\rangle =\widehat{P}\vert \Psi_0\rangle=\prod_i \widehat{P_i}\vert \Psi_0\rangle,
\end{equation}
with
\begin{equation} \label{eq:P}
\widehat{P}_i=\sum_\Gamma \lambda_{i\Gamma} \widehat{m}_{i\Gamma},
\end{equation}

\begin{equation}\label{eq:m_I_projection}
\widehat{m}_{i\Gamma}=\left\vert i,\Gamma \right\rangle \left\langle i,\Gamma \right\vert,
\end{equation}

where $\widehat{m}_{i\Gamma}$ is the projector to the specified configuration $\vert \Gamma \rangle$
on site \textit{i}. In Eq. (\ref{eq:psi}), the role of projection operator $\widehat{P}$ is to
adjust the weight of each atomic configuration through variational parameters $\lambda_{i\Gamma}(0\le
\lambda_{i\Gamma} \le 1)$. The GWF falls back to KSWF if all $\lambda_{i\Gamma}=1$. On the other hand,
if $\lambda_{i\Gamma}=0$, the configuration $\vert \Gamma \rangle$ on site \textit{i} will be totally
removed. In this way, both the itinerant behavior of uncorrelated wave functions and the localized
behavior of atomic configurations can be described consistently, and the GWF can give a more accurate
description of the correlated metallic systems than KSWF.

The total energy of the above system can be expressed as the expectation value of the Hamiltonian equation
(\ref{eq:H}) using GWF, which takes the form

\begin{equation} \label{eq:energy}
E_{Total} = \langle \Psi _{G}|H|\Psi _{G}\rangle = \langle \Psi _{G}|H_{LDA}|\Psi _{G}\rangle
+\langle \Psi _{G}|H_{int}|\Psi _{G}\rangle-E_{DC},
\end{equation}

In Eq. (\ref{eq:energy}), the interaction energy is given as
\begin{equation} \label{eq:HGint}
  \langle\Psi_{G}\vert H_{int}\vert\Psi_{G}\rangle=\sum_{i,\Gamma}E_{i\Gamma}m_{i\Gamma},
\end{equation}
where $m_{i\Gamma}$ is the weight of configuration $\Gamma$,
\begin{equation}
  m_{i\Gamma}= \langle\Psi_{G}\vert \widehat{m}_{i\Gamma} \vert \Psi_{G} \rangle
\end{equation}

According to Eq. (\ref{eq:psi}), the LDA energy of Eq. (\ref{eq:energy}) can be written as

\begin{equation} \label{eq:ELDA}
  \langle \Psi _{G}|H_{LDA}|\Psi _{G}\rangle=\langle \Psi _{0}| \widehat{P} H_{LDA} \widehat{P} |\Psi _{0}\rangle
  = \langle \Psi _{0}| H_{LDA}^{G} |\Psi _{0}\rangle.
\end{equation}

 $H_{LDA}^{G}$ is called the effective Hamiltonian under Gutzwiller approximation.

The DFT calculations for realistic materials are always done in reciprocal space, so the formulas above
should transform to the reciprocal space. We define the Bloch states of localized orbitals $\vert i\alpha
\rangle$

\begin{equation} \label{eq:alpha}
\vert k\alpha\rangle=\frac{1}{N}\sum_{i}e^{ikR_{i}}\vert i\alpha \rangle
\end{equation}

Then $H_{LDA}^{G}$ in \textit{k} space can be written as

 \begin{equation} \label{eq:HG}
\begin{split}  H_{LDA}^{G}
 & =\left( \sum_{k\alpha}z_{\alpha}\vert k\alpha\rangle\langle k\alpha\vert+1-\sum_{k\alpha}\vert k\alpha\rangle\langle k\alpha\vert \right) \\
 & \quad H_{LDA} \left( \sum_{k'\beta}z_{\beta}\vert k'\beta\rangle\langle k'\beta\vert+1-\sum_{k'\beta}\vert k'\beta\rangle\langle k'\beta\vert \right) \\
 & +\sum_{kk'\alpha} \left(1-z_{\alpha}^{2} \right)\vert k\alpha\rangle\langle k'\alpha\vert H_{LDA}\vert k'\alpha\rangle\langle k\alpha\vert,\end{split}
\end{equation}

where $z_{\alpha}$ is the renormalization factor for local orbital $\alpha$, which depends on those Gutzwiller
variational parameters $\lambda_{\Gamma}$; for those noninteracting orbitals, the corresponding $z$
factor equals 1.

According to Eqs. (\ref{eq:energy}), (\ref{eq:HGint}), (\ref{eq:ELDA}), and (\ref{eq:HG}), the total energy reads

\begin{equation} \label{eq:total}
\begin{split}  E_{Total}
 & =\langle \Psi_{0}\vert \left( \sum_{k\alpha}z_{\alpha}\vert k\alpha\rangle\langle k\alpha\vert+1-\sum_{k\alpha}\vert k\alpha\rangle\langle k\alpha\vert \right) \\
 & \quad H_{LDA} \left( \sum_{k'\beta}z_{\beta}\vert k'\beta\rangle\langle k'\beta\vert+1-\sum_{k'\beta}\vert k'\beta\rangle\langle k'\beta\vert \right) \vert\Psi_{0}\rangle \\
 & +\sum_{\alpha} \left( 1-z_{\alpha}^{2} \right) n_{\alpha} \varepsilon_{LDA}^{\alpha} +\sum_{\Gamma}E_{\Gamma}m_{\Gamma}-E_{DC},\end{split}
\end{equation}

where $\varepsilon_{LDA}^{\alpha}=\sum_{k}\langle k \alpha \vert H_{LDA} \vert k \alpha \rangle$ and
$n_{\alpha}=\sum_{k}\langle \Psi_{0}|k \alpha \rangle \langle k \alpha | \Psi_{0} \rangle$.

The total energy expressed in Eq. (\ref{eq:total}) depends on both the uncorrelated ``starting''
wave function $\vert \Psi_{0}\rangle$ and those Gutzwiller variational parameters $\lambda_{\Gamma}$,
which can both be determined by minimizing the total energy. After we obtain the ground-state wave
function, we can calculate most of the ground-state properties based on it; please refer to our paper\cite{xydeng_2009} for more details.

\section{results and discussions} \label{sec:results}

\begin{figure}[!ht]
\includegraphics[width=10cm]{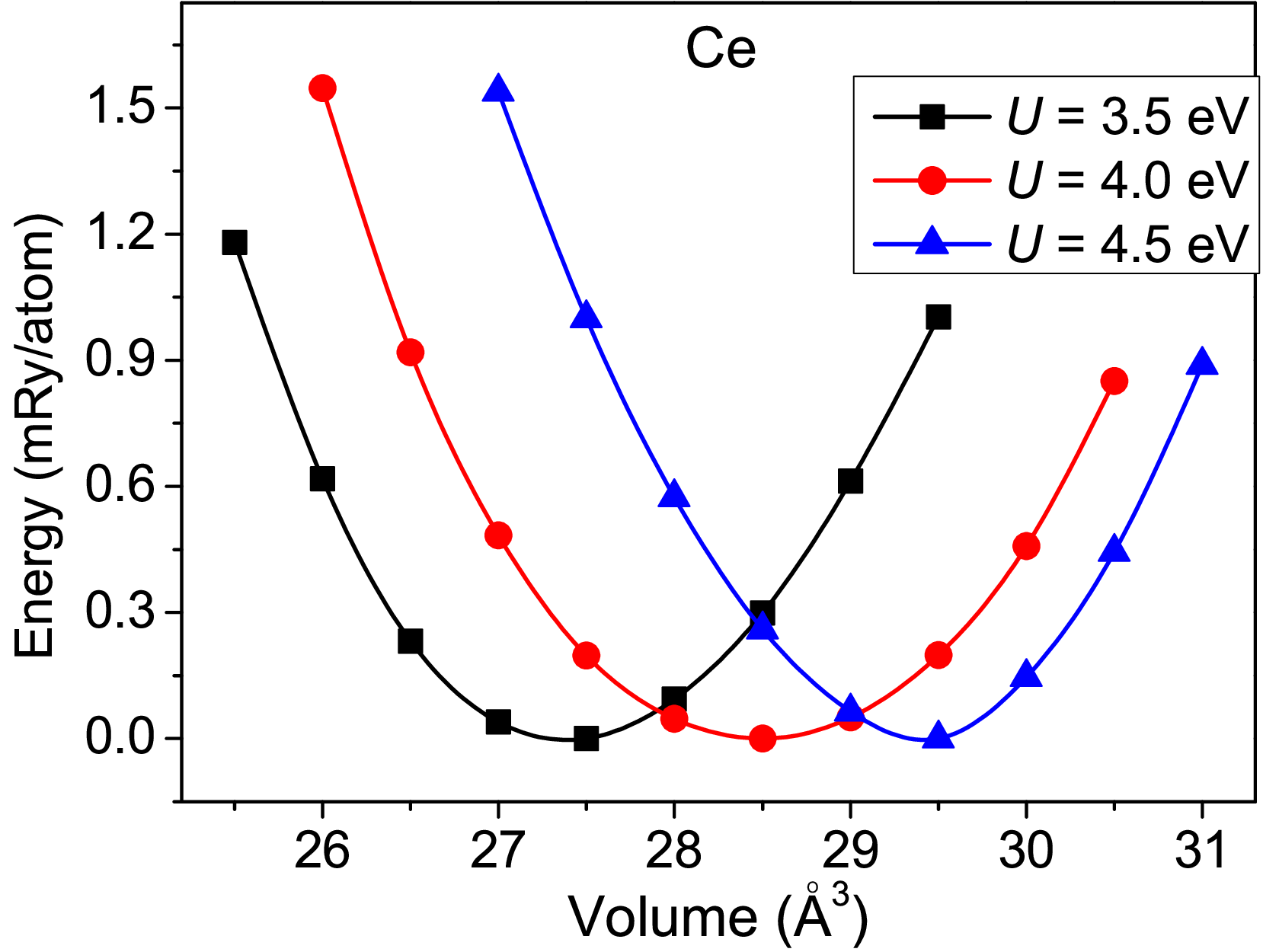}
\caption{(Color online) Calculated energy curves of $\alpha$-Ce (fcc) as a function of atomic volume for
different values of $U$ by LDA+G method.} \label{ev}
\end{figure}

Like LDA+U and LDA+DMFT methods, in LDA+G the Hubbard-like local Coulomb interaction \textit{U} will
be chosen as the only empirical parameter. First we calculate the equilibrium volume
for $\alpha$-Ce (fcc) to check the validity of the different \textit{U} value as shown in Fig. \ref{ev}.
The volume of Ce under ambient pressure is between 28.0 and 29.0 \AA$^3$ reported experimentally,\cite{Ellinger_1974, Olsen_1985} so the equilibrium volume for $U=4.0$ eV (28.49 \AA$^3$)
is in good agreement with experiments, while the equilibrium volume for $U=3.5$ eV (around 27.5 \AA$^3$)
and $U=4.5$ eV (around 29.5 \AA$^3$) are smaller and larger than
experiments, respectively. We also calculate the bulk modulus for $\alpha$-Ce under ambient pressure
as shown in Table \ref{tab:bulk} together with the results from the all-electron FPLMTO calculations,\cite{Soderlind_1995a} pseudopotential plane-wave calculations,\cite{Richard_2001} and
experiments.\cite{Ellinger_1974, Olsen_1985} We can see from the table that the LDA calculation usually
over-estimates the bonding strength among cerium atoms, which makes the equilibrium volume obtained
by LDA to be about 20\% smaller and bulk modules to be much larger than experiments. After treating
the correlation effect more carefully by the Gutzwiller variational method, our results are in good
agreement with experiments. Thus we set $U=4.0$ eV for all the calculations for cerium with different
volume, which is consistent with other LDA+U and DMFT calculations, and spin-orbital coupling effect
is always fully included.

\begin{table}[H]
\caption{Theoretical and experimental values of the equilibrium volume \textit{V} and bulk modulus \textit{B}
for $\alpha$-Ce from our LDA+G calculations, some LDA/GGA calculations (Refs. \cite{Richard_2001} and
\cite{Soderlind_1995a}), and experimental data (Refs. \cite{Ellinger_1974} and \cite{Olsen_1985}).}
\begin{tabular}{ c @{\hspace{1cm}} c @{\hspace{1cm}} c }
\hline
\hline
        & \textit{V} (\AA$^3$)  &  \textit{B} (GPa) \\
\hline
LDA     & 23.3\cite{Richard_2001}, 22.74\cite{Soderlind_1995a} & 58.7\cite{Richard_2001}, 60.5\cite{Soderlind_1995a} \\
GGA     & 26.3\cite{Richard_2001}, 26.05\cite{Soderlind_1995a} & 43.0\cite{Richard_2001}, 48.7\cite{Soderlind_1995a} \\
Experiment\cite{Ellinger_1974, Olsen_1985}     &  28-29  &  20-35  \\
Present (LDA+G) &  28.49  &  27.6  \\
\hline
\hline
\end{tabular}\label{tab:bulk}
\end{table}

\begin{figure}[!ht]
\includegraphics[width=10cm]{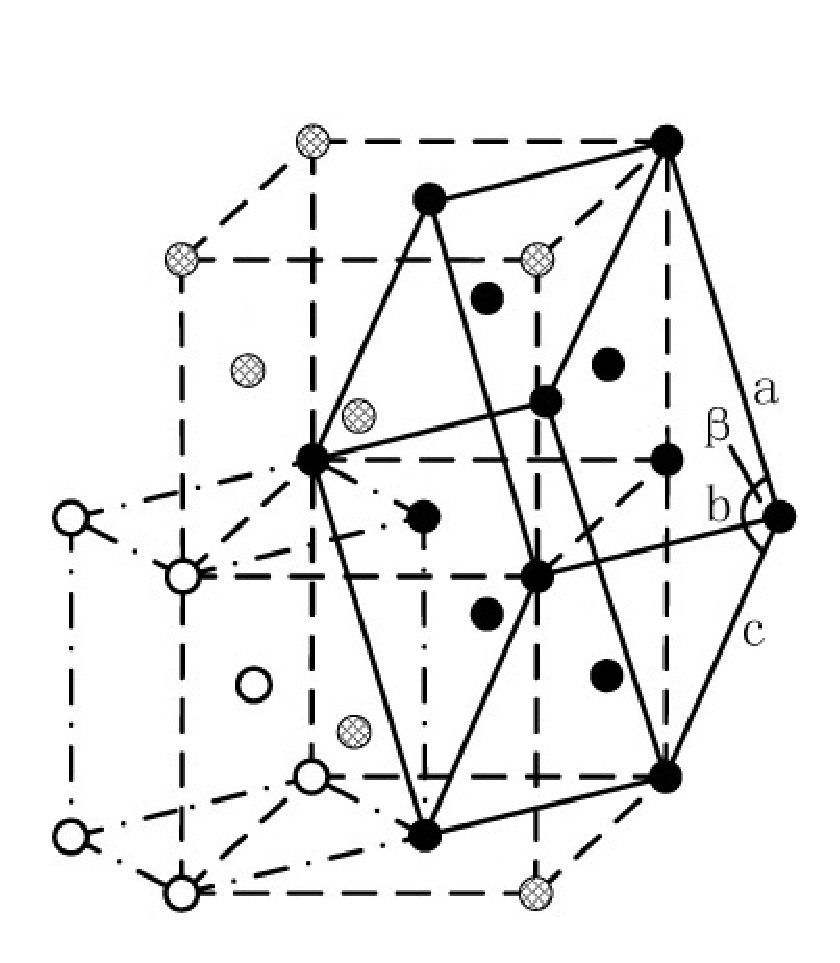}
\caption{The relationship between the $\alpha$ phase (face-centered cubic, dashed
lines with hatched circles---main cell), bct phase (body-centered tetragonal,
dash-dot line subcell with unfilled circles), and the $\alpha''$ phase (C-face-centered monoclinic,
solid line subcell with black circles).} \label{lattice}
\end{figure}

\begin{figure}[!ht]
\includegraphics[width=10cm]{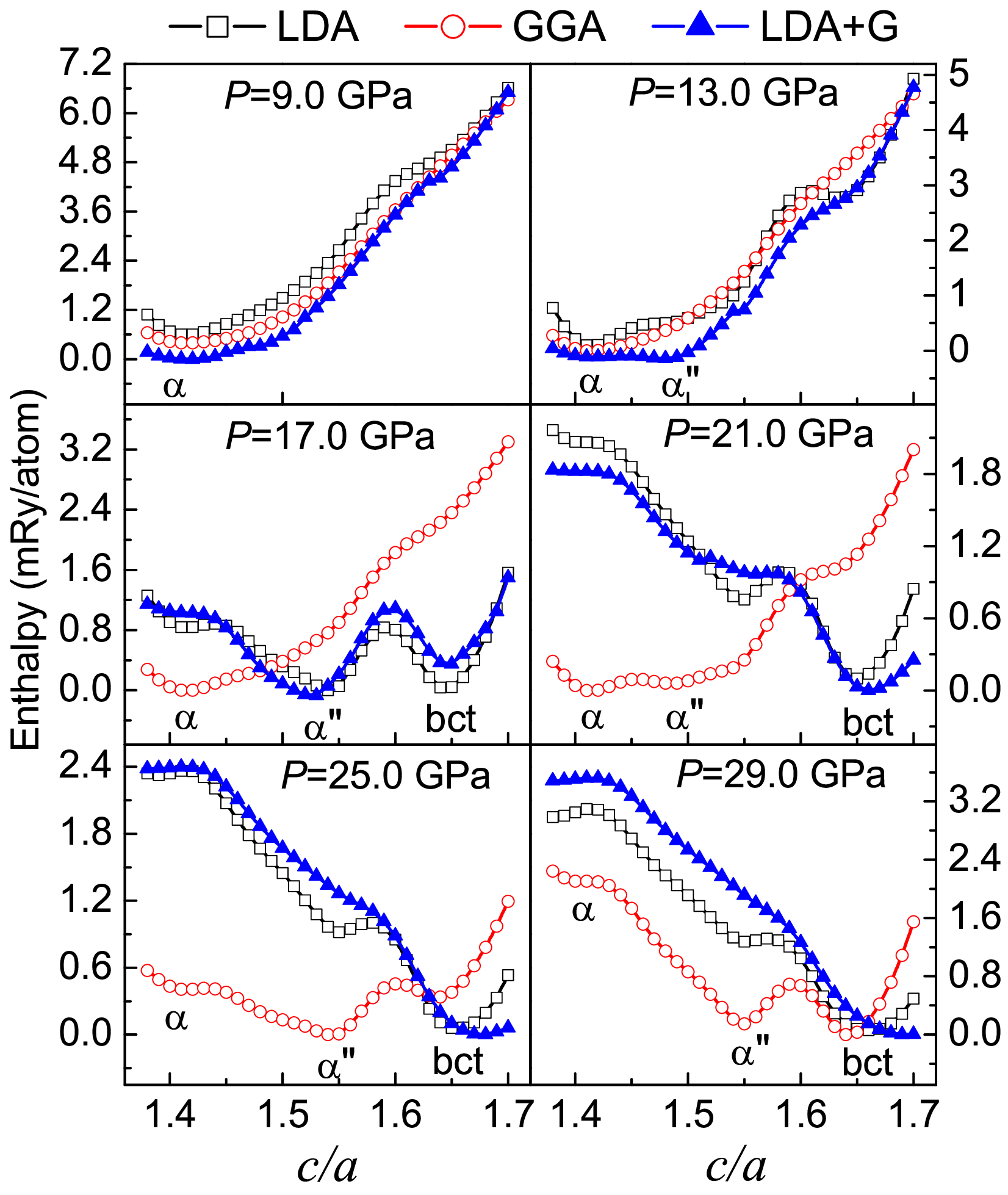}
\caption{(Color online) Calculated enthalpy of body-centered-tetragonal Ce as a function of
$c/a$ axial ratio by LDA, GGA, and LDA+G methods, respectively.}
\label{bainpath}
\end{figure}

Early experiments have reported two intermediate-pressure phases of cerium, $\alpha'$ phase and $\alpha''$
phase, together with the low-pressure $\alpha$ phase and high-pressure bct phase. The thermodynamic
stability of these four phases under pressure is the main interest of the present work. Neglecting the
tiny distortion in the $\alpha''$ phase, all these three phases including $\alpha$, $\alpha''$, and the
high-pressure bct phase can be treated within the same bct structure but with the different ratio
of the lattice constants $c/a$,\cite{Olsen_1985, McMahon_1997, Soderlind_1995b} which is illustrated in Fig. \ref{lattice}. The $c/a$ ratio is exactly $\sqrt2$ for $\alpha$ phase with the fcc structure and is found to
be around 1.65 for the high-pressure bct phase. The $c/a$ ratio of the $\alpha''$ phase is reported experimentally
to be around $1.5 \le c/a \le 1.56$.\cite{Ellinger_1974, Olsen_1985} Therefore in the present paper,
we first apply the LDA+G method to minimize the enthalpy of the cerium with bct structure with respect
to the $c/a$ ratio as the function of pressure, which mimics the competition among the $\alpha$, $\alpha''$,
and high-pressure bct phases under pressure, and after that we will compare the enthalpy of these
phases and the $\alpha'$ phase. Our main results have been plotted in Fig. \ref{bainpath} with the
comparison to LDA and GGA. The results obtained by all three methods agree quite well for pressure
less than 9.0 GPa indicating that the $\alpha$ phase (fcc) with $c/a=\sqrt2$ is the thermodynamic
stable structure. For pressure between 9.0 and 25.0
GPa, the results obtained by LDA, GGA, and LDA+G are quite different. First of all, all three methods
predict that all phases appear as the locally stable phases in this intermediate pressure
region, while LDA results indicate that the $\alpha''$ phase
is the thermodynamic stable phase in a very small region around 17.0 GPa. Although GGA gives a reasonably
wide region from 23.0 to 27.0 GPa, within which the $\alpha''$ phase is globally stable, this pressure
region is much higher than the experimental results, which is from 6.9 to 12.0 GPa.\cite{Olsen_1985}
LDA+G predicts that the $\alpha''$ phase is the thermodynamic stable phase in the pressure region
of 13.0\textit{--}17.0 GPa, which is much closer to the experimental data.\cite{Olsen_1985} For
pressure larger than 25.0 GPa, all three methods again reach the same conclusion that the high-pressure
bct phase is thermodynamic stable, whereas the other two are either metastable or unstable.

\begin{figure}[!ht]
\includegraphics[width=10cm]{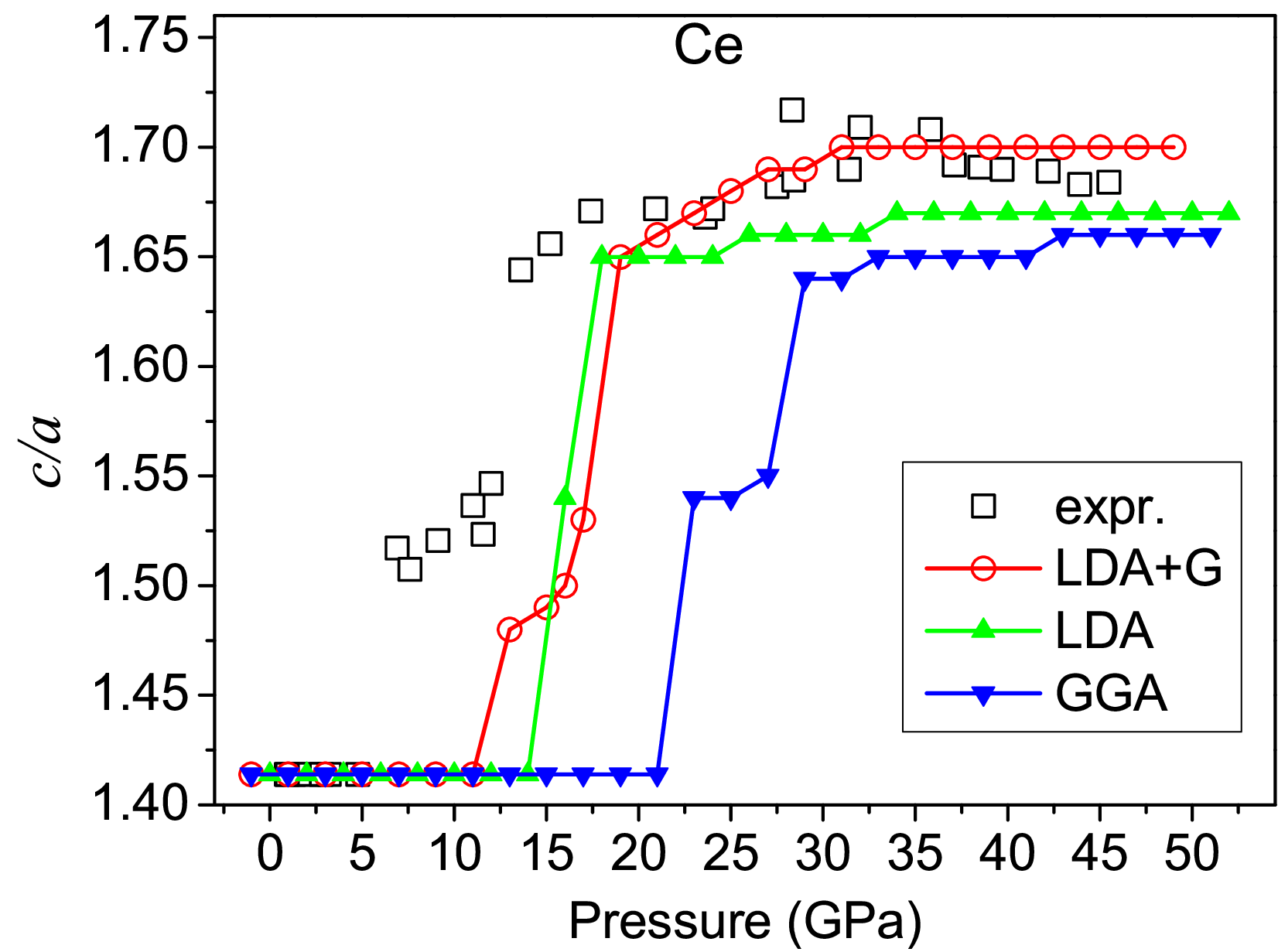}
\caption{(Color online) The $c/a$ axial ratio for the body-centered tetragonal structure as a function of
pressure for Ce. Experimental data (Ref. \cite{Olsen_1985}) are marked with black open squares, while LDA+G
results are given by a red solid line and open circles. LDA results are given by a green solid line
and filled uptriangles, and the results of GGA are shown by a blue solid line and filled downtriangles.} \label{expr_comp}
\end{figure}

The optimized $c/a$ ratio as the function of pressure for Ce obtained from our LDA, GGA, and LDA+G calculations
are plotted in Fig. \ref{expr_comp} together with the experimental results.\cite{Olsen_1985} We can
find clearly that the LDA+G calculation obtains the globally stable region of $\alpha''$-Ce to
be from  13.0 to 17.0 GPa, which is much closer to the experimental data,\cite{Olsen_1985}
while LDA only gets a very narrow region for $\alpha''$-Ce stable, and GGA gets the $\alpha''$ phase
stable in a pressure region that is much higher than experimental data, as shown in Fig. \ref{expr_comp}.

\begin{figure}[!ht]
\includegraphics[width=9cm]{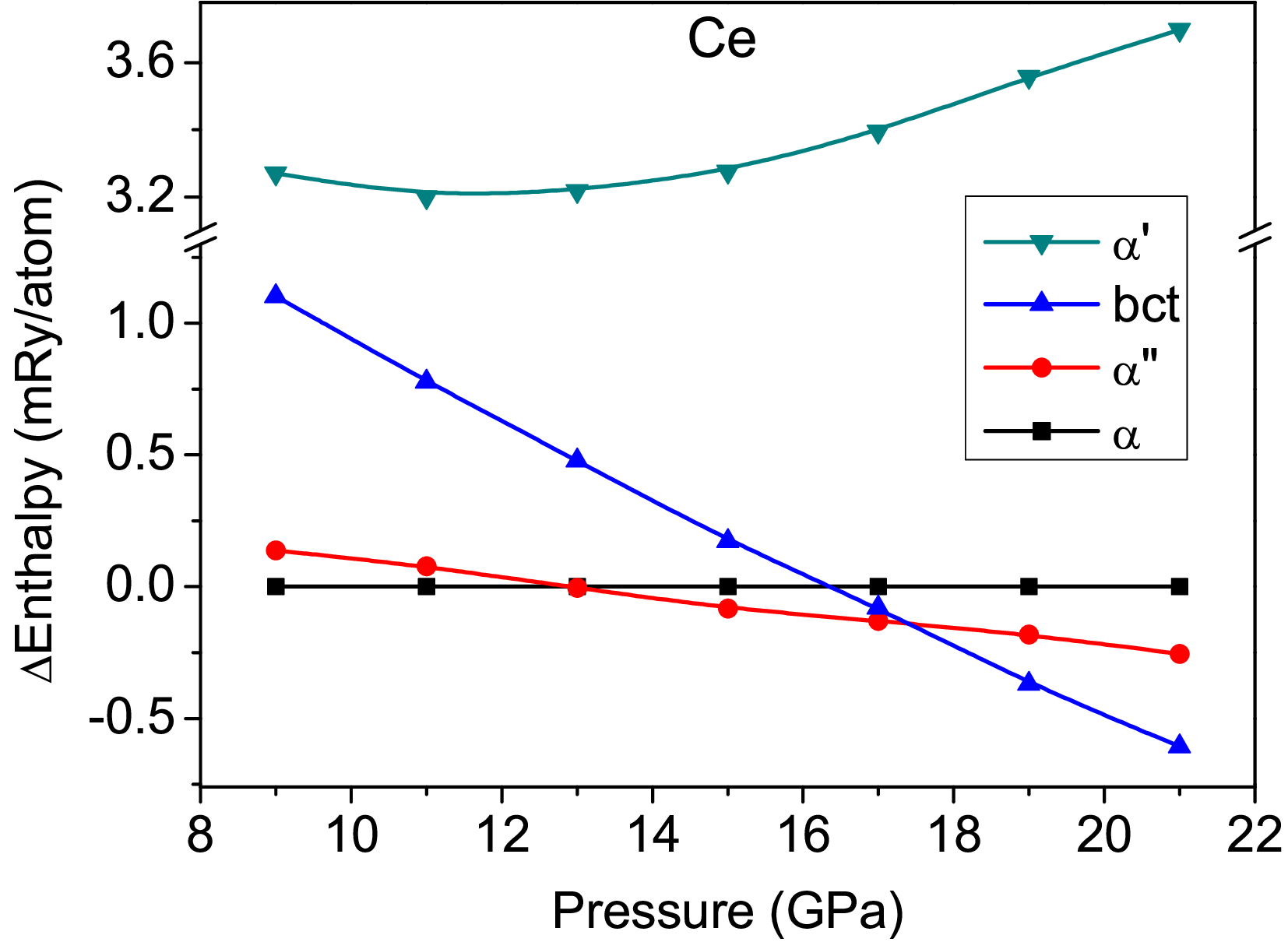}
\caption{(Color online) The enthalpy curves for the intermediate-pressure phases of Ce relative to the $\alpha$ phase (fcc) 
as obtained from LDA+G calculations.} \label{mul-phase}
\end{figure}

The next issue to be addressed is the relative enthalpy of the $\alpha'$ phase compared with the other
three phases discussed in the previous paragraph. The $\alpha'$ phase has a orthorhombic $\alpha$-U
structure, which can be viewed as distorted fcc with some of the face-centered atoms being shifted from
their original positions, as described by the parameter $2y$.\cite{Skriver_1985} The $2y$ value obtained
experimentally by McMahon and Nelmes\cite{McMahon_1997} is 0.2028 \AA. If $2y=0.5$ \AA and $a=b=c$,
the standard fcc structure can be restored. Therefore we can calculate the energy of $\alpha$-Ce (fcc)
and $\alpha'$-Ce within the same $\alpha$-U structure. We use $a/b=0.5115, c/b=0.8756$ as obtained from
the experiment,\cite{McMahon_1997} and optimize the $2y$ value for any given volume. We find that $2y=0.21$
gives the minimum energy for volume being 22.5 \AA$^3$, which is in  good agreement
with experiment.\cite{McMahon_1997} We calculated the enthalpy of $\alpha$-Ce (fcc) and $\alpha'$-Ce within
the same $\alpha$-U structure frame
for a pressure region: 9.0 GPa $\le P \le$ 21.0 GPa. The enthalpy difference between them ($H_{\alpha'}-H_{\alpha}$) is plotted in Fig. \ref{mul-phase} together with the enthalpy difference of $H_{\alpha''}-H_{\alpha}$
and $H_{bct}-H_{\alpha}$ obtained previously. Our results confirm that the $\alpha'$ phase is always higher
in enthalpy in the entire pressure region considered in the present paper.
We thus rule out the $\alpha'$ structure as a thermodynamic stable intermediate
pressure phase of cerium. This conclusion is in good agreement with the FPLMTO  calculations.\cite{Ravindran_1998}
From Fig. \ref{mul-phase} we can also see that the $\alpha''$ structure is the thermodynamic stable phase
among the above-mentioned four possible phases within the pressure region 13.0\textit{--}17.0 GPa.
Therefore, based on the LDA+G calculation, we conclude that the $\alpha''$ phase is the thermodynamic stable
phase for cerium in the intermediate pressure region. This conclusion is quite consistent with the most
recent experiments.\cite{McMahon_1997, Dmitriev_2004}

\begin{figure}[!ht]
\includegraphics[width=10cm]{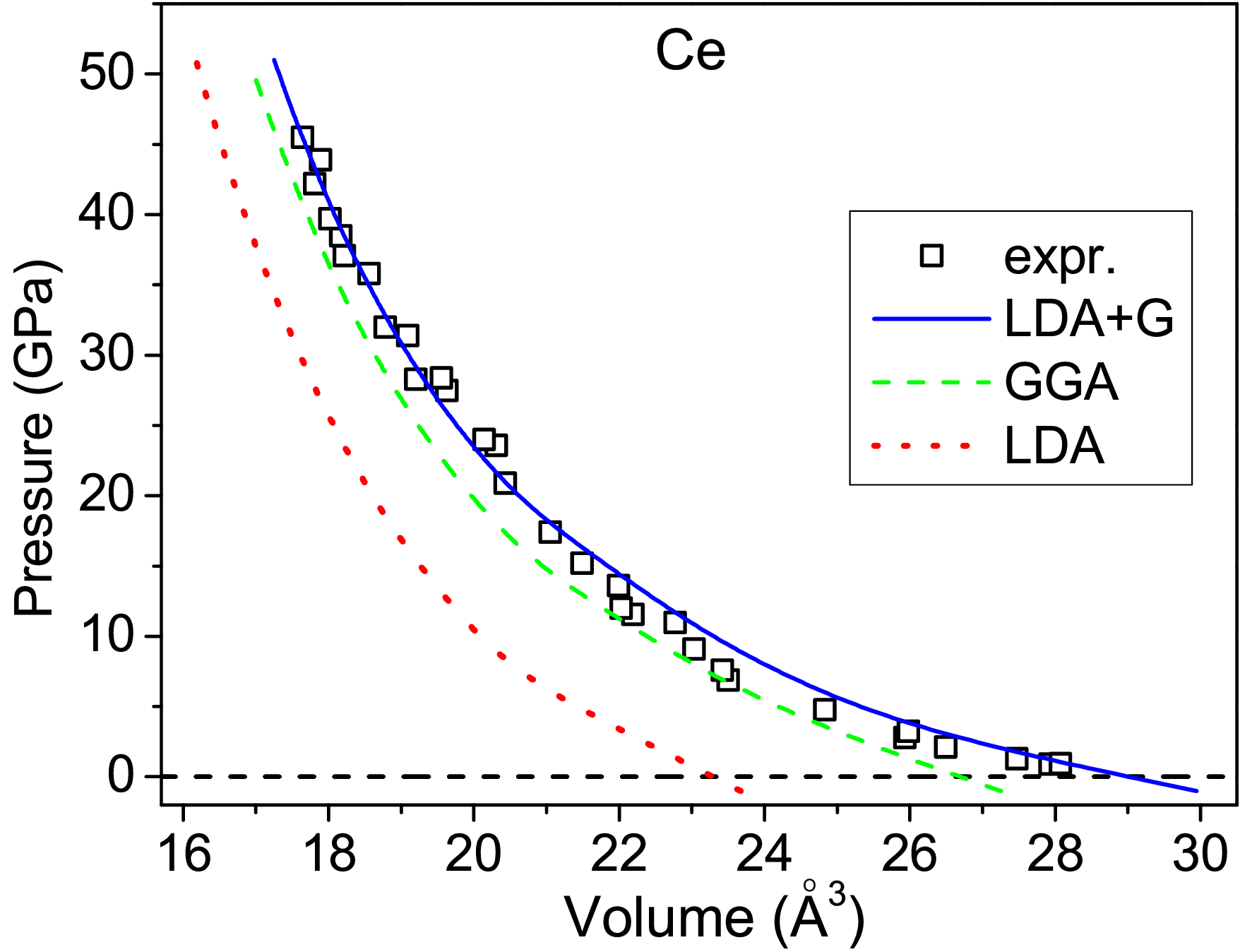}
\caption{(Color online) Equation of state for Ce. Experimental results (Ref. \cite{Olsen_1985}) are marked with open squares and theory is given by lines. } \label{eos}
\end{figure}

Based on the calculations in the previous paragraph, we obtain the multiphase equations of state (EOSs)
for cerium metal, as shown in Fig. \ref{eos} together with the experimental data.
The agreement between LDA+G results and the experimental data\cite{Olsen_1985} is very
good, while the GGA curve is slightly away especially in the low-pressure region and
 the LDA result is further away from the experimental data.

\begin{figure}[!ht]
\includegraphics[width=10cm]{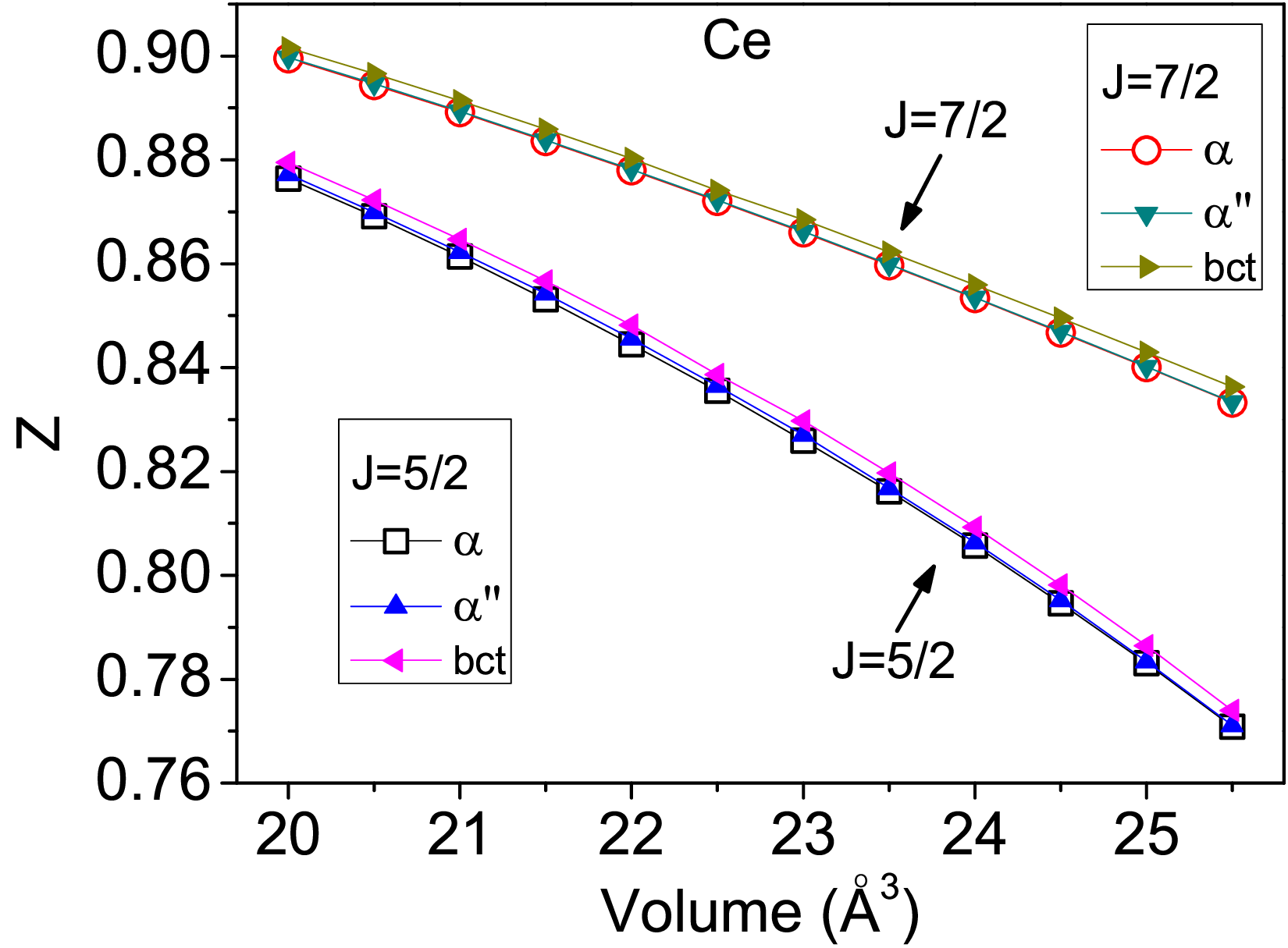}
\caption{(Color online) The average renormalization factor as a function of volume for angular momentum $J=5/2$ 
and $J=7/2$ \textit{f} electrons of Ce.} \label{zfac}
\end{figure}

In Fig. \ref{zfac}, we plot the renormalization factor of the 4\textit{f} bands in the $\alpha$, $\alpha''$,
and high-pressure bct phases as the function of volume. We can find that the renormalization factor of the
4\textit{f} bands decreases monotonically with the increment of the volume for all the three phases,
which can be easily explained by the fact that increasing volume reduces the hopping integral between
the neighboring \textit{f} orbitals, which enhances the correlation effect among 4\textit{f} electrons
and thus reduces the corresponding renormalization factor. From the present LDA+G calculation, we find
that the main consequence of the correlation effect in the total energy is to reduce the kinetic energy.
Since the fcc structure is close packed, compared with the $\alpha''$ phase, the $\alpha$ phase has
relatively higher kinetic energy gain, which is overcounted by LDA. Because of that, the reduction
of kinetic energy gain captured by LDA+G is also more pronounced in the $\alpha$ phase, which raises
the total energy of the $\alpha$ phase relative to the $\alpha''$ phase and makes it thermal dynamically
unstable in the intermediate pressure region. We would like to emphasize that the renormalization factor 
obtained in Gutzwiller approximation is defined by the reduction of the kinetic energy, which is in general
higher than that obtained by dynamical mean-field theory through the quasiparticle spectral weight of
the Green's function. Please refer to our previous paper for the detailed discussion on this point.\cite{xydeng_2008}

\section{conclusions} \label{sec:conclusions}

In summary, using the newly developed LDA+G technique, we have carried out systematical numerical study
on the phase diagram of cerium metal under pressure. We found that the correlation effect among \textit{f}
electrons in cerium plays a crucial role to determine the thermodynamic stable phase of cerium in the intermediate
pressure region. The LDA calculation overestimates the chemical bonding contributed by the 4\textit{f}
electrons, which leads to smaller equilibrium volume and larger bulk modulus compared to the experimental
data. With the increment of pressure, the overlap between 4\textit{f} orbitals becomes more and more
pronounced, which reduces the correction to the total energy caused by the correlation effect. Therefore
the correct description of the correlation effect, which evolutes with the pressure, becomes one of the
key issues to obtain the correct phase diagram in the intermediate pressure region. Our numerical results
obtained by the LDA+G method conclude that the $\alpha''$ phase is the thermodynamic stable phase in the
intermediate-pressure region, which is quite consistent with the recent experiments.

ACKNOWLEDGMENTS: The authors thank L. Wang, J. N. Zhuang, G. Xu, and Q. M. Liu for their helpful
discussions. We acknowledge the support from the 973 Program of China (Grant No. 2007CB925000), from the
NSF of China (Grants No. NSFC 10876042 and No. NSFC 10874158), and that from the Development Foundation
of CAEP (Grant No. 2008A0101001).

%1

%2

%3

%4

%5

%6

%7

\end{document}